\documentclass[12pt,preprint]{aastex}

\shorttitle{IFS of 1RXS J1131-1231}
\shortauthors{Sugai et al.}

\begin{document}
\defcitealias{slu03}{S03}
\defcitealias{mor06}{M06}
\defcitealias{bla06}{BPR06}

\title{Integral Field Spectroscopy of the Quadruply Lensed Quasar
1RXS~J1131-1231: New Light on Lens Substructures
\altaffilmark{1}}

\author{H. Sugai\altaffilmark{2},
        A. Kawai\altaffilmark{2},
        A. Shimono\altaffilmark{2},
        T. Hattori\altaffilmark{3,4},
        G. Kosugi\altaffilmark{5,6},
        N. Kashikawa\altaffilmark{7},
        K. T. Inoue\altaffilmark{8},
        and 
        M. Chiba\altaffilmark{9}}

\altaffiltext{1}{Based on data collected at Subaru Telescope,
which is operated by the National Astronomical Observatory of Japan.}
\altaffiltext{2}{Department of Astronomy,
Kyoto University, Sakyo-ku, Kyoto 606-8502, Japan. \\
email: sugai@kusastro.kyoto-u.ac.jp}
\altaffiltext{3}{Okayama Astrophysical Observatory,
National Astronomical Observatory of Japan,
Kamogata-cho, Asakuchi-gun, Okayama 719-0232, Japan}
\altaffiltext{4}{Present Address: Subaru Telescope,
National Astronomical Observatory of Japan,
650 North A'ohoku Place, Hilo, Hawai'i 96720, U.S.A.}
\altaffiltext{5}{Subaru Telescope,
National Astronomical Observatory of Japan,
650 North A'ohoku Place, Hilo, Hawai'i 96720, U.S.A.}
\altaffiltext{6}{Present Address: ALMA Project,
National Astronomical Observatory of Japan,
Mitaka, Tokyo 181-8588, Japan}
\altaffiltext{7}{Optical and Infrared Astronomy Division,
National Astronomical Observatory
of Japan,
Mitaka, Tokyo 181-8588, Japan}
\altaffiltext{8}{School of Science and Engineering, Kinki University,
Higashi Osaka 577-8502, Japan}
\altaffiltext{9}{Astronomical Institute,
Tohoku University, Sendai 980-8578, Japan}

%%%%%%%%%%%%%%%%%%%%%%%%%%%%%%%%%%%%%%%%%%%%%%%%%%%%%%%%%%
\begin{abstract}
We have observed the quadruply lensed quasar 1RXS J1131-1231 with
the integral field spectrograph mode of the Kyoto Tridimensional
Spectrograph II mounted on the Subaru telescope. Its field of view
has covered simultaneously 
the
three brighter lensed images A, B, and C,
which are known to exhibit anomalous flux ratios in their continuum
emission. We have found that the [OIII] line flux ratios among these
lensed images are consistent with those predicted by smooth-lens
models. The absence of both microlensing and millilensing effects
on this [OIII] narrow line region sets important
limits on the mass of any substructures along the line of sight, 
which is expressed as $M_{\rm E} < 10^5 M_\odot$ for
the mass inside an Einstein radius.
In contrast, the H$\beta$ line emission, which originates from
the broad line region, shows an anomaly in the flux ratio between
images B and C, i.e., a factor two smaller 
C$/$B ratio
than predicted by smooth-lens models.
The ratio of A$/$B in the H$\beta$ line is well reproduced. 
We show that the anomalous C$/$B ratio for the 
H$\beta$ line is caused most likely by micro/milli-lensing of image C.
This is because other effects, such as the differential dust
extinction and/or arrival time difference between images B and C,
or the simultaneous lensing of another pair of images A and B,
are all unlikely. In addition, we have found that the broad H$\beta$
line of image A shows a slight asymmetry in its profile compared
with those in the other images, which suggests the presence of a small
microlensing effect on this line emitting region of image A.
\end{abstract}
%%%%%%%%%%%%%%%%%%%%%%%%%%%%%%%%%%%%%%%%%%%%%%%%%%%%%%%%%%

\keywords{dark matter --- gravitational lensing --- quasars: emission lines
 --- quasars: individual (1RXS J1131-1231)}

\section{INTRODUCTION}
\label{intro}

Gravitational lensing is an important probe for the mass distribution of
lensing galaxies as well as for the internal structure of lensed sources
\citep[e.g.,][]{koc06}.
Quadruply lensed quasars
are particularly useful in this regard since the positions and
flux ratios of the lensed images provide strong constraints on lens models,
and provide far more observational information than doubly imaged systems.
For instance, recently revealed quads with ``anomalous flux ratios'', namely
flux ratios which are unexplained in smooth-lens models, suggest the presence
of numerous substructures in the lensing galaxy
\citep[e.g.,][]{mao98,met01,chi02,dal02,sch02,kee03}.
Whether or not these substructures correspond to cold dark matter (CDM)
subhalos (milli-lensing), stellar populations (micro-lensing), or differential
reddening by dust, remains an issue, and depends on the lens system
\citep[e.g.,][]{chi05}.
It is of particular interest to assess whether there is a high level of substructures or
subhalos in galaxy-sized halos, as predicted in CDM models
\citep{kly99,moo99}.

A technique to distinguish the origin of anomalous flux ratios
was proposed by 
\citet{mou03},
utilizing spectroscopic
information of lensed quasars. If we take into account a finite (i.e.,
not point-like) source size of a quasar heart, which consists of a small
continuum emitting region (CR) with typical size $R_{\rm S}$ of
$\sim 10^{-4}$~pc, broad line region (BLR) with $R_{\rm S}\sim 1$~pc,
and narrow line region (NLR) with $R_{\rm S} \ga 100$~pc,
then the lensing magnification will be functions of these source sizes and
the mass of any substructure inside its Einstein angle, $M_{\rm E}$
(i.e., either CDM subhalos with a mass $M_{\rm E} \sim10^{7-9} M_\odot$ or
stars with $M_{\rm E} \sim 1 M_\odot$). If CDM subhalos are responsible for
anomalous flux ratios, then they can magnify both the CR and BLR because
their Einstein radii (projected onto the distance of the quasar),
$r_{\rm E} \sim 0.01 (M_{\rm E}/M_\odot)^{1/2} h^{-1/2}$~pc (where
$h=H_0/100$ km~s$^{-1}$~Mpc$^{-1}$),
are larger than the sizes of the CR and BLR. The NLR may be magnified as well,
depending on the upper mass of the CDM subhalos. On the other hand,
stars would be unable to magnify emission lines from either the BLR or NLR, while
still magnifying the CR. As a consequence, depending on the nature of substructure,
the fluxes of emission lines in each of the lensed images that show anomalous
(continuum) flux ratios can be either magnified or unchanged.

The observations of multiply lensed objects
with an integral field spectrograph (IFS) is useful
in this study, since it provides simultaneous information among quasar
images both spatially and spectrally \citep[e.g.,][]{met04,way05}.
The IFS is 
more superior to conventional slit spectroscopic techniques
in the following aspects:
(1) Slit spectroscopic observations would require separate observations
for each pair of components. This not only requires more observing time
but also introduces significant errors and uncertainties for relative
flux measurements since exposures for each pair are carried out
in different
observational conditions.
(2) IFS allows us to optimize the relative flux measurements
{\it after} the observations, in terms of determining the
size and center positions of the apertures used for photometry.

In this paper, we report on our spectroscopic mapping of the quadruply
imaged quasar, 1RXS J1131$-$1231, which shows anomalous flux ratios, using 
the IFS mode of Kyoto tridimensional spectrograph II
\citep[Kyoto 3DII:][]{sug00b,sug02,sug04} mounted on the Subaru 8.2-m telescope.
This lens system is nearly unique among known gravitationally lensed systems
in that it brings together rare properties, including the quadruplicity,
a bright optical Einstein ring, a small redshift, and high amplification
\citep[hereafter S03]{slu03}. \citetalias{slu03} measured redshifts of
0.658 and 0.295 respectively for the quasar and the lensing galaxy.
The lens shows three roughly co-linear highly magnified images,
A, B, and, C. This configuration emerges if the quasar is close to and
inside a cusp point of the astroid caustic for an elliptical lens
(see also Figure~\ref{oiiiextensionmodel} shown in \S 4).
In such a lens system associated with a cusp singularity, there exists a
universal relation between the image fluxes, $(F_B+F_C)/F_A=1$
\citep[e.g.,][]{mao98}, whereas the observed flux ratios violate these rules
significantly [$(F_B+F_C)/F_A \simeq 2.1$ in the $V$ band and $2.2$ in the $R$ band
\citepalias{slu03}]. The flux ratios appear to be changing with time,
as a result of microlensing effects on some of the images. Time delay effects
between the images do not play a role in the flux ratio anomaly because
simple smooth-lens
models predict only a day or a fraction of a day for the time delays between
images A, B, and C \citep{slu06,cla06}.
Several different lens models with smooth mass distribution have been
constructed to attempt to reproduce the image positions, flux ratios, and
time delays of this lens system. But it appears that one requires a high-order
lens structure in addition to a simple ellipsoidal lens, in the form of
an octupole component or an unidentified satellite 
\citep[][hereafter M06]{cla06,mor06}.
 
Armed with the IFS
mode of Kyoto 3DII,
we measured the emission-line fluxes of
both the BLR H$\beta$ and the NLR [OIII]$\lambda\lambda$4959,5007 for images
A, B, and C simultaneously. The measurements of the line fluxes are more
reliable than those of the continuum because there are no contributions
from the lensing galaxy, the redshift of which differs from the quasar's.
The H$\beta$ and [OIII] lines are very close in wavelength, so that
the effect of differential reddening between them is almost totally negligible.
The unique fine sampling of
$\sim 0^{\prime\prime}.1$, together with the simple shape of the point spread
function and Subaru's excellent image quality, provides the most suitable
method for measuring the relative line fluxes.

This paper is organized as follows.
\S 2 describes the observations and method for data reduction.
Spectra of images A, B, and C and flux ratios between each image pair are
shown in \S 3. \S 4 is devoted to discussion and conclusions.

\section{OBSERVATIONS AND REDUCTION}
\label{obs}

%% In a manner similar to \objectname authors can provide links to dataset
%% hosted at participating data centers via the \dataset{} command.  The
%% second curly bracket argument is printed in the text while the first
%% parentheses argument serves as the valid data set identifier.  Large
%% lists of data set are best provided in a table (see Table 3 for an example).
%% Valid data set identifiers should be obtained from the data center that
%% is currently hosting the data.

We observed
1RXS J1131-1231 on 2005 February 8
with the IFS mode of
the visitor instrument Kyoto 3DII, mounted on the Subaru.
We used the atmospheric dispersion corrector installed at
the Subaru Cassegrain focus \citep{iye04}.
This enabled us to obtain the atmospheric-dispersion-free
data cube without having to apply a correction during the data reduction
\citep[e.g.,][]{sug05}.
The IFS mode uses an array of $37 \times 37$ lenslets, enabling us
to obtain
spectra of
$\sim 10^3$
spatial elements.
The spectral range from
$7300 \AA$ to $9150 \AA$
was observed 
in each of two one-hour exposures.
With the spatial sampling of $0^{\prime\prime}.096$ lenslet$^{-1}$
\citep{sug04},
the field of view
of $\sim 3^{\prime\prime}$
covered
the three brighter
quasar images.
The target was located at the same position of the lenslet array
during the two exposures.
Since the positional difference derived from the reconstructed data 
cubes was small, within a few tenths of a lenslet (i.e., a few
$\times 0^{\prime\prime}.01$),
we combined them without any offset between them.
The spatial resolution was stable during the observations, and was
determined to be
$0^{\prime\prime}.5$-$0^{\prime\prime}.6$
from a strong ^^ ^^ point"-like H$\beta$ emission-line
distribution in the quasar image A (see \S~\ref{spectabc}).

We used
halogen lamp spectral frames 
for the flat fielding.
These frames
were obtained immediately before and after
each of the target frames, with the same optical setting as the
target frames.
Since our calibration optics exactly simulates the telescope optics,
i.e, since it provides the same micropupil images as the telescope
does, it properly corrects differences in the lenslet-to-lenslet 
response (and efficiency).
The uniformity of a flat-fielded reconstructed image is $1.5\%$
(1 $\sigma$).
The wavelength calibration was carried out  using He and Ne
lamps, whose spectra were
also
taken 
immediately before and after
each of
the target
frames. 
We measured that the calibration lamp line
positions changed only by 0.1 pixel ($\sim 0.3 \AA$)
along the spectral direction during the whole sequence of observations.
This corresponds to a systematic uncertainty in the wavelength
calibration.
Moreover,
the technique of micropupil spectroscopy
\citep{sug06}
provides us
with an accuracy in the relative wavelength calibration of
$\sim 0.2 \AA$, which has been determined within each of the
calibration lamp frames.
The full width at half maximum of the instrumental line profile
was measured as $7 \AA$, which
corresponds to 260 km s$^{-1}$.
In order to improve signal-to-noise ratios in target spectra,
we smoothed them in wavelength by convolution with a
gaussian profile
so that the resultant
velocity resolution was 390 km s$^{-1}$.
The sky subtraction was carried out using
simultaneous sky spectra.
This is particularly important for longer wavelengths,
where the sky emission lines are strong. 
For an accurate sky subtraction the Kyoto 3DII has two
separate fields of view:
one for the target and the other smaller one for sky,
which is located $33^{\prime\prime}$ away from the target
\citep{sug06}.
The averaged sky spectrum was subtracted from each
target lenslet spectrum
(see \S~\ref{spectabc}).

The IFS data of the spectroscopic standard star HD93521
were used 
for the instrument response curve correction
as well as 
for 
absolute
flux
density
calibration.
The extinction curve at Mauna Kea
(CFHT Observatory Manual,
http://www.cfht.hawaii.edu/Instruments/
ObservatoryManual/)
was used to correct 
for the effects of airmass in both the target object and standard star spectra.
The airmasses were 1.38 and 1.22 for the target frames and
1.23 for the standard star frames. 
The atmospheric conditions were checked using a guide star count
log and were determined to have been non-photometric.
We roughly estimate the uncertainty of absolute flux density
calibration to be about $\pm 20\%$.

\section{SPECTRA OF QUASAR IMAGES A, B, AND C}
\label{spectabc}

Figure~\ref{lineimage} shows line images and a line-free
continuum image that have been obtained from the
IFS data.
We find an extended component of the [OIII] emission connecting
quasar images A and C.
Such an extended component is not seen
between quasar images A and B.
The spatial distribution
of the line-free continuum emission is similar to that of
the H$\beta$ emission rather than the [OIII] emission.
Despite the severe contamination from the host galaxy
(\S~\ref{compab}),
the continuum is much fainter in image C than in
image B.
Figure~\ref{abcab} (upper)
shows the spectra of quasar images A, B, and C that have each been
extracted with
an 8-lenslet, i.e.
a circular aperture with a diameter $0^{\prime\prime}.77$.
An aperture of this size includes $\sim 60$ \% of the total flux of a point source.
The aperture centers have been determined from the H$\beta$
image shown in Figure~\ref{lineimage}.
The relative positions of these centers are consistent,
to within 
several
milliarcseconds, with peak positions
obtained by \citetalias{mor06} using the
{\it Hubble Space Telescope}.
As an example, the lower panel in Figure~\ref{abcab} shows the spectrum of image A 
both with and without sky subtraction
in order to demonstrate the effectiveness of
the sky subtraction as described in \S~\ref{obs}.

\subsection{Comparison Between A and B}
\label{compab}

If all the emission from the quasar is magnified in the same manner
within each quasar image, 
we should be able to completely subtract the spectrum of one quasar
image with the spectrum of another quasar image by
scaling 
the latter with an appropriate factor. 
If this is not the case,
e.g. when the NLR line is magnified only by
a macrolens while the BLR line is further magnified by a 
milli/micro-lens,
it should be 
impossible to subtract these two lines completely with a single
multiplicative factor.
Using such spectral fittings, we derive the flux ratios in the BLR 
and in the NLR below.
We do not attempt to derive the flux ratios in the continuum, which
is severely contaminated by the quasar host galaxy.

Figure~\ref{aboiiihbbroad} (upper)
shows 
a comparison between quasar images A \& B.
The spectrum of image A has been fit, using the least squares method,
in the [OIII] and the line-free continuum emission 
wavelength regions
according to the following equation:
$f_A(\lambda) = b_0 \times f_B(\lambda) -
b_1 \times (\lambda/8000 \AA)^{b_2}$,
where $f_A(\lambda)$ and $f_B(\lambda)$ are the flux densities
of images A and B, respectively.
The fitting parameter $b_0$ determines the scaling factor,
while $b_1$ and $b_2$ are used to remove the smooth continuum.
The spectrum has been fit with $b_0=1.63$.
Similarly, we have fit
the spectrum of image A 
in the broad H$\beta$ and the continuum
wavelength regions
(Figure~\ref{aboiiihbbroad} (lower)).
The broad H$\beta$ in image A
seems to have only slightly weaker
emission in the bluer wing,
compared with that in the scaled image B.
The best fit $b_0$ is 1.74.
We therefore conclude that
the flux ratios A$/$B in the [OIII] and broad H$\beta$ emission lines
are $1.63^{+0.04}_{-0.02}$ and
$1.74^{+0.07}_{-0.12}$ respectively, 
where the uncertainties are derived as discussed 
in \S~\ref{uncertainty}.
In order to compare these values with the ratio predicted
by macrolens models,
we have constructed a smooth-lens model with a singular
isothermal ellipsoid plus an external shear
\citep{kas93,kor94},
based on the quasar image positions obtained by \citetalias{mor06}.
A slight offset of the halo center with respect to the 
lensing 
galaxy center has been 
allowed.
This model predicts an A$/$B ratio of 1.66.
The smooth-lens models constructed by \citetalias{slu03} and
by \citet[][hereafter BPR06]{bla06}, with a singular
isothermal sphere plus an external shear,
also predicts a similar ratio of 1.75 - 1.70.
We conclude therefore that the measured ratios are close
to those predicted by smooth-lens models.
The required offset
for the spectral fitting,
i.e. non-zero $b_1$,
is likely caused primarily by
contamination from the quasar
host galaxy continuum.
The contamination from the lensing galaxy is small
since it
is located outside of our field of view:
$\sim 2^{\prime\prime}.1$ \citepalias[e.g.,][]{mor06}
down from image A in Figure~\ref{lineimage}.
Figure~\ref{lineimage} (right) suggests that
the contribution of the lensing galaxy is less than
$6 \times 10^{-18}$ erg cm$^{-2}$ s$^{-1}$ $\AA^{-1}$
and actually much smaller
since we do not detect any signature of it even
at positions that are closer to its center. 

\subsection{Comparison Between B and C}
\label{compbc}

We have compared the spectra of images B and C
using the same method as in \S~\ref{compab}.
When we scale 
the spectra
according to the [OIII] line emission
(Figure~\ref{bcoiiihbbroad} (upper)),
we find differences in the H$\beta$.
The H$\beta$ profile also differs in these two spectra.
Figure~\ref{bcoiiihbbroad} (lower)
shows the same comparison but with the C spectrum scaled
according to the broad H$\beta$ emission.
We find differences in the [OIII] lines and in the narrow
line (NL) H$\beta$.
The redshift of the 
NL
H$\beta$ matches that of the 
[OIII] line 
($z_{heliocentric}=0.6542 \pm 0.0001 (1 \sigma)$,
where the uncertainty includes the accuracy of [OIII]
line peak determination as well as the wavelength calibration
uncertainties).
The 
^^ ^^ NL-subtracted" 
spectrum in
Figure~\ref{bcoiiihbbroad} (upper)
suggests that the
broad line (BL) H$\beta$ is more symmetric than the
original 
BL$+$NL
H$\beta$,
and is redshifted with respect to the 
NL
H$\beta$ ($z_{heliocentric}=0.6581 \pm 0.0001$ for BL H$\beta$).
We have measured an asymmetry parameter, $A$, of the H$\beta$
emission
for both the original and residual spectrum of image B:
$A=(HWHM_{Red} - HWHM_{Blue})/FWHM$,
where $HWHM_{Red}$ and $HWHM_{Blue}$ represent the half-widths at
half-maximum
flux density of the red and blue
wings of the profile respectively \citep{wil93}.
Our measurements have shown that A=0.32 and A=0.08 respectively,
for the original BL$+$NL H$\beta$ and the residual BL H$\beta$.
The difference in the H$\beta$ profile between images B and C
is caused by different fractions of NL
contribution.
The above comparisons lead us to conclude that
the flux ratio between images B and C largely depends on
where the lines originate: the BLR or the NLR.
The flux ratio C$/$B in the [OIII] line is $1.19^{+0.03}_{-0.12}$,
whereas it is $0.46^{+0.02}_{-0.03}$ in the BL H$\beta$ line.
As before, uncertainties are derived as in \S~\ref{uncertainty}.
Compared to 
predictions of
the smooth-lens models,
this is larger only
by a factor of $\sim 1.2$
in [OIII]
but is smaller by a factor of $\sim 2.1$
in the 
BL
H$\beta$.
Table~\ref{tbl-ifsspect} summarizes the flux ratios
obtained between the quasar images A, B, and C,
together with
the ratios expected from the smooth-lens models.

In order to derive the H$\beta$ and [OIII] flux ratios
between quasar images,
we have used the simple scaling and fitting methods above,
without using any template for the FeII line emission.
This is because we have found that the 
observed
FeII line ratios
differ slightly from the well-established I~Zw~1
template \citep{ver04}
and that the FeII
is better subtracted
by the scaling between quasar images themselves.

Figure~\ref{psfmoffat} demonstrates in another way
the suitability of our scaling method.
The H$\beta$ residual image has been created by subtracting
three point sources
with
the H$\beta$
flux ratios derived above, from the
H$\beta$
image
shown in Figure~\ref{lineimage}.
The residuals have an rms of only $3 \%$
of the peak value of 
quasar image A. 

\subsection{Uncertainty Estimates}
\label{uncertainty}

The uncertainties of the measured flux ratios in
Table~\ref{tbl-ifsspect} have been estimated as follows.
First we have tried another spectrum fitting procedure that differs
from the procedure used in Figures~\ref{aboiiihbbroad} and
\ref{bcoiiihbbroad}.
In this new fitting procedure, we fit the line emission after
removing the continuum emission from each image spectrum,
rather than fitting the line and continuum emission simultaneously.
This altered the 
flux ratios
slightly, 
by 0.00--0.05.
Secondly we have estimated the
mutual flux contamination
between images
due to point-like components, 
by using the model of three point sources
shown in Figure~\ref{psfmoffat}.
In the $0^{\prime\prime}.77$ aperture the contamination in broad H$\beta$ for
images A, B, and C are 1 \%, 3 \%, and 5 \% respectively.
The effects on the A$/$B and C$/$B ratios are $-2 \ \%$ and $+2 \ \%$.
In the same aperture the mutual contaminations between the
^^ ^^ point-like" [OIII] components 
(see \S~\ref{discon}) are estimated as
2 \%, 3 \%, and 2 \% respectively.
The effects on the A$/$B and C$/$B ratios are $-1 \%$.
Thirdly
we have derived the flux ratios with the simultaneous
(line $+$ continuum)
fitting procedure, but using successively smaller apertures, 
down to the size of a 4-lenslet ($0^{\prime\prime}.38$ diameter).
While the C$/$B ratio in [OIII] decreases
by up to 0.07 as the aperture size is reduced,
the other ratios vary by no more than $\pm 0.02$.
These differences should in practice include various effects:
mutual contamination,
contributions from the spatially extended 
components,
readout and photon noise, and flat
fielding uncertainties.
Lastly we have considered the uncertainty of the A$/$B ratio in
the broad H$\beta$ emission due to the slight difference in its
wing profile between images A and B (\S~\ref{compab}).
When, during the fitting, we reduce the weighting of each
blue wing data point by half, the A$/$B ratio is 1.76.
When we reduce the weighting of each red wing data 
point by half, on the other hand, the A$/$B ratio is 1.69.
We have taken all the above into consideration when determining the 
uncertainties of the measured flux ratios in Table~\ref{tbl-ifsspect}.
The effects of uncertainties of the aperture center positions
are negligible because the positional uncertainties
are smaller than one tenth of a lenslet.

\section{DISCUSSION AND CONCLUSIONS}
\label{discon}

Anomalous flux ratios in lensed images can be induced by substructure in a host
lens. As a reference for investigating such substructure lensing, we estimate
the angular size of the radius of a line-emitting region, $\theta_{\rm S}$, and
of an Einstein radius for a lens substructure, $\theta_{\rm E}$,
based on the set of cosmological parameters $\Omega=0.3$, $\Lambda=0.7$,
and $h=0.7$. A source image with radius $R_{\rm S}$ at the redshift 0.658 of
the quasar has an angular size of
$\theta_{\rm S} \simeq 1.4 \times 10^{-4} (R_{\rm S}/1 {\rm pc})$ arcsec.
For comparison, a lens having a mass $M_{\rm E}$ inside an Einstein angle yields
$\theta_{\rm E} \simeq 6.7 \times 10^{-7} (M_{\rm E}/0.1 M_\odot)^{1/2}$
arcsec for the lens redshift of 0.295. Thus, a dimensionless ratio
$\xi \equiv \theta_{\rm S} / \theta_{\rm E}$ showing the degree of a lensing
effect \citepalias{bla06} is given as
$\xi \simeq 200 (R_{\rm S}/1 {\rm pc})(M_{\rm E}/0.1 M_\odot)^{-1/2}$.
\citet{wyi02} showed, based on detailed lensing simulations, that
$\theta_{\rm E}$ must be at least an order of magnitude smaller than
$\theta_{\rm S}$ for no magnification. This leads us to require that $\xi \ga 10$
if there is no additional magnification of an image by a lens substructure.
Otherwise an image should be subject to flux anomaly.

As presented in Table~\ref{tbl-ifsspect}, the flux ratios for the [OIII]
line emission, which originates from the NLR with $R_{\rm S} \ga 100$ pc,
are basically in agreement with those of a standard smooth lens consisting
of an elliptical lens and an external shear.
Also, the [OIII] flux ratios appear to be inconsistent with 
\citetalias{mor06}'s
lens
model, which posits a massive satellite with $\sim 5 \times 10^{10}
M_\odot$ near image A in order to reproduce their time delay measurements.
Such a massive lens perturber would affect the [OIII] flux significantly.
The basic agreement of the [OIII] flux ratios with the smooth-lens model
predictions suggests not only the validity of the smooth-lens
models but also the absence of millilensing effects.
A star with subsolar mass is unable to magnify the NLR because
$\xi \sim 2 \times 10^4$ for $R_{\rm S}=100$ pc and $M_{\rm E}=0.1$ $M_\odot$.
A more massive substructure such as a CDM subhalo, even if it exists
near the light path to the NLR, is tightly constrained to have
$M_{\rm E} < 10^5$ $M_\odot$.

\begin{figure}
\epsscale{1.0}
\plotone{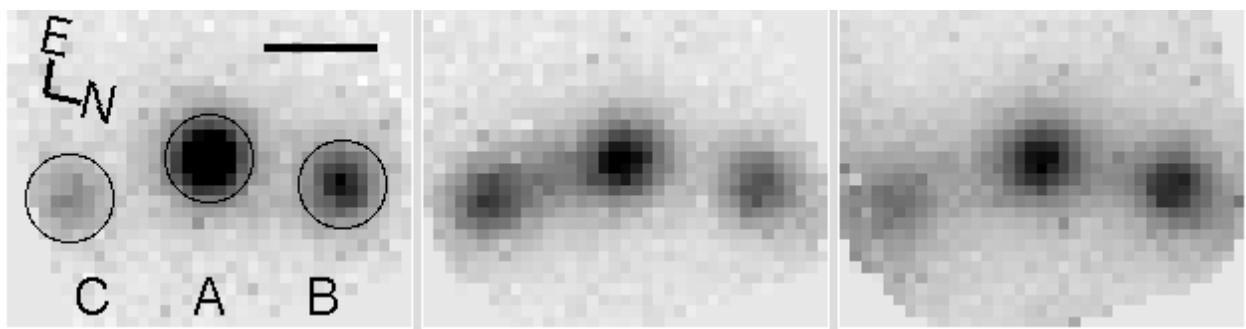}
\caption{
Images extracted from our IFS data.
{\it (left)}
Continuum-subtracted H$\beta$(+FeII$\lambda$4924)
line image. The spectra have been
integrated for the observed wavelength region from
$7947 \AA$ to $8162 \AA$.
The continuum level has been determined in
wavelength regions
from $7878 \AA$ to $7947 \AA$ and
from $8399 \AA$ to $8469 \AA$.
The length of a bar to the upper right
is $1^{\prime\prime}$.
The apertures used for obtaining spectra for quasar images,
such as spectra in Figure~\ref{abcab}, are shown
with
circles.
North is at $102^{\circ}$ clockwise from the top.
{\it (middle)}
Continuum(+FeII$\lambda$5018)-subtracted
[OIII]$\lambda$5007 line image.
The spectra have been
integrated for the observed wavelength region from
$8247 \AA$ to $8300 \AA$.
The ^^ ^^ continuum" level has been determined in
wavelength regions from $8221 \AA$ to $8247 \AA$ and
from $8300 \AA$ to $8343 \AA$.
{\it (right)}
Line-free continuum image.
The spectra have been
integrated for the observed wavelength regions
from $7878 \AA$ to $7947 \AA$ and
from $8399 \AA$ to $8469 \AA$.   
\label{lineimage}
}
\end{figure}

\begin{figure}
\epsscale{0.8}
\plotone{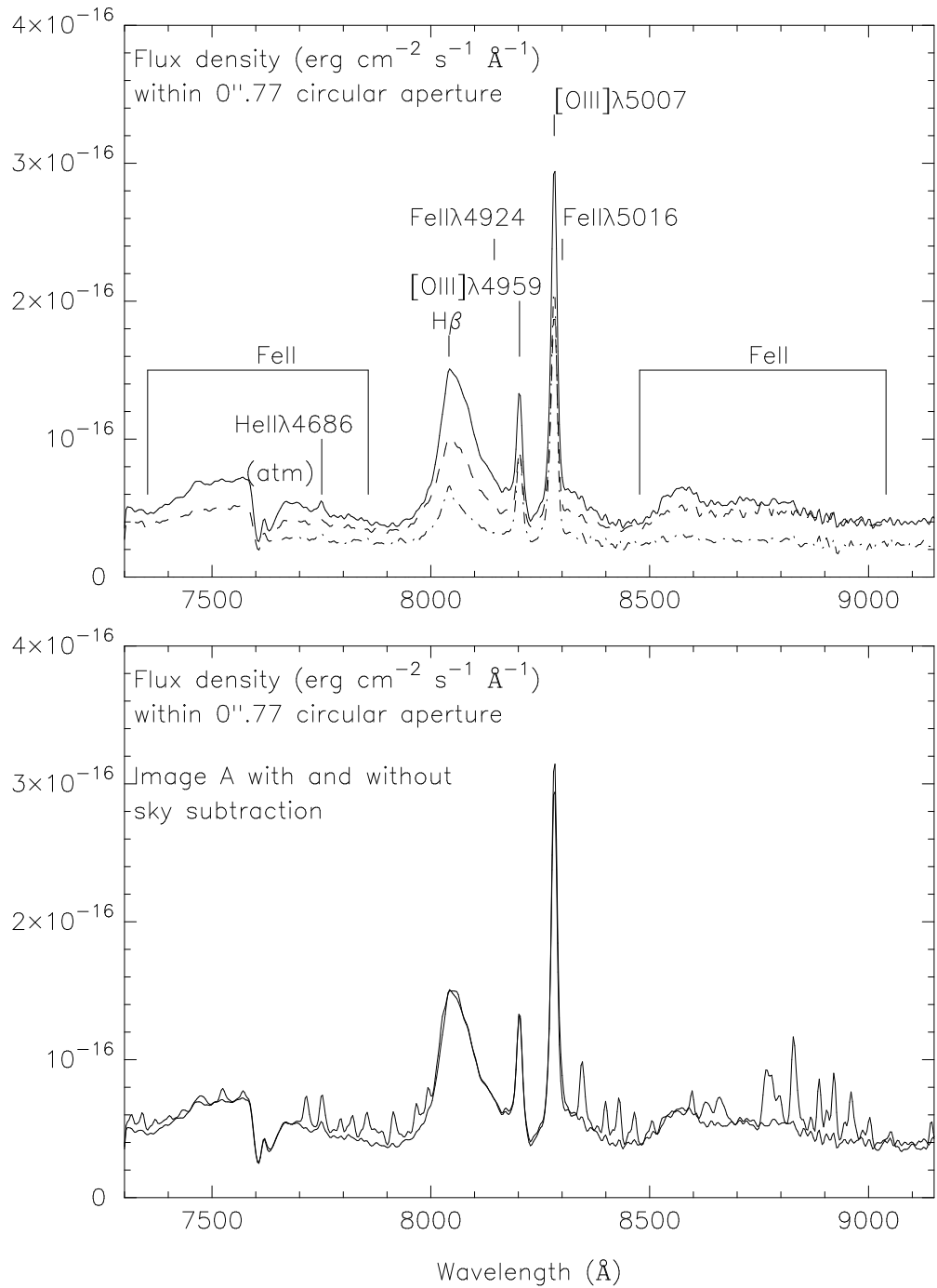}
\caption{
{\it (upper)}
Spectra of quasar images A(solid line), B(dashed line),
and C(dot-dashed line) that have been extracted with
an 8-lenslet ($0^{\prime\prime}.77$) pure circular aperture each.
The apertures are shown in
Figure~\ref{lineimage}.
The spectra, as well as those in Figures~\ref{aboiiihbbroad}
and \ref{bcoiiihbbroad}, have not been corrected in flux
density due to the finite aperture.
The shorter wavelength part of image C is not shown,
because 
image C 
was located
close to 
an edge of the lenslet array
and its shorter wavelength part was outside of the 
CCD 
frame
when the micropupil images corresponding to image C
were dispersed by a grism.
We have carefully removed such a wavelength region so that the
image C spectrum shown does not suffer from 
such flux losses due to the grism dispersion.
Similarly, the longer wavelength part of image B is not shown.
{\it (lower)}
Spectrum of image A is shown for both cases with and without
sky subtraction.
\label{abcab}
}
\end{figure}

\begin{figure}
\epsscale{0.7}
\plotone{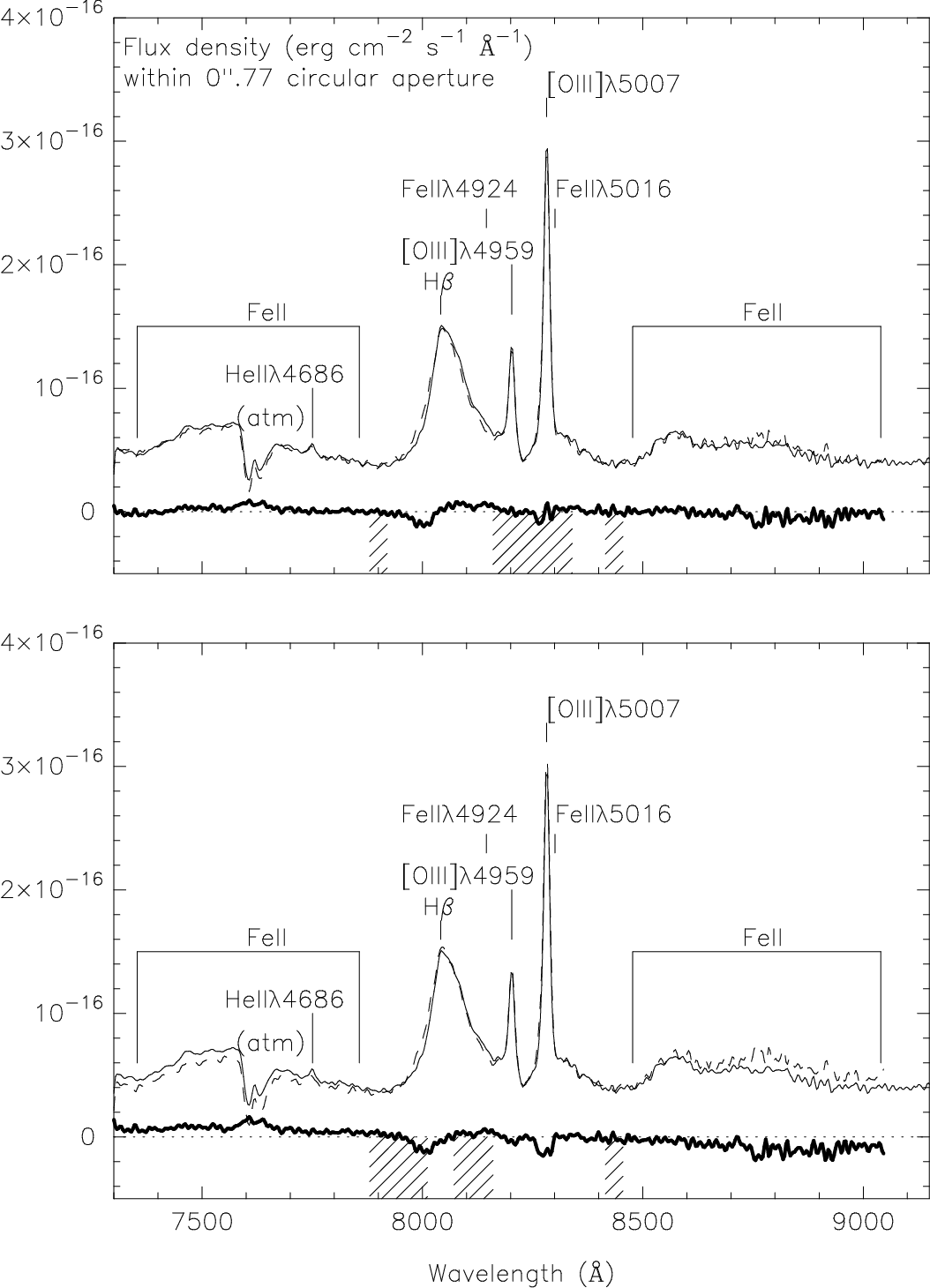}
\caption{
{\it (upper)}
The spectrum of image A (thin solid line) has been fit with
the following equation:
$f_A(\lambda) = b_0 \times f_B(\lambda) -
b_1 \times (\lambda/8000 \AA)^{b_2}$,
where $f_A(\lambda)$ and $f_B(\lambda)$ are the flux densities
of images A and B, respectively,
and $b_0$, $b_1$, and $b_2$ are the fitting parameters.
The wavelength regions used for the fitting,
$7880$-$7920 \AA$, $8160$-$8340 \AA$,
and $8415$-$8455 \AA$, are shown with shadows.
These correspond to
the [OIII] and the line-free continuum emission 
wavelength regions.
The dashed line is the best fit
($b_0=1.63$, $b_1=1.69 \times 10^{-17}$, and $b_2=1.9$),
while the thick solid line shows the residuals,
$f_A(\lambda) -$(best fit).
The value of $b_2$ is rather meaningless since the wavelength
ranges used for the fitting are too narrow to give its accurate
value.
This causes apparent residuals in the FeII regions, which are
the outsides of the fitting wavelength regions. 
{\it (lower)}
The same as
the above,
except
the wavelength regions used for the fitting
and the resultant fitting parameters.
The wavelength regions are
$7880$-$8011 \AA$, $8071$-$8160 \AA$,
and $8415$-$8455 \AA$,
which correspond to
the broad H$\beta$ and the line-free continuum emission 
wavelength regions.
The best fit parameters are
$b_0=1.74$, $b_1=2.26 \times 10^{-17}$, and $b_2=-1.6$.
\label{aboiiihbbroad}
}
\end{figure}

\begin{figure}
\epsscale{0.7}
\plotone{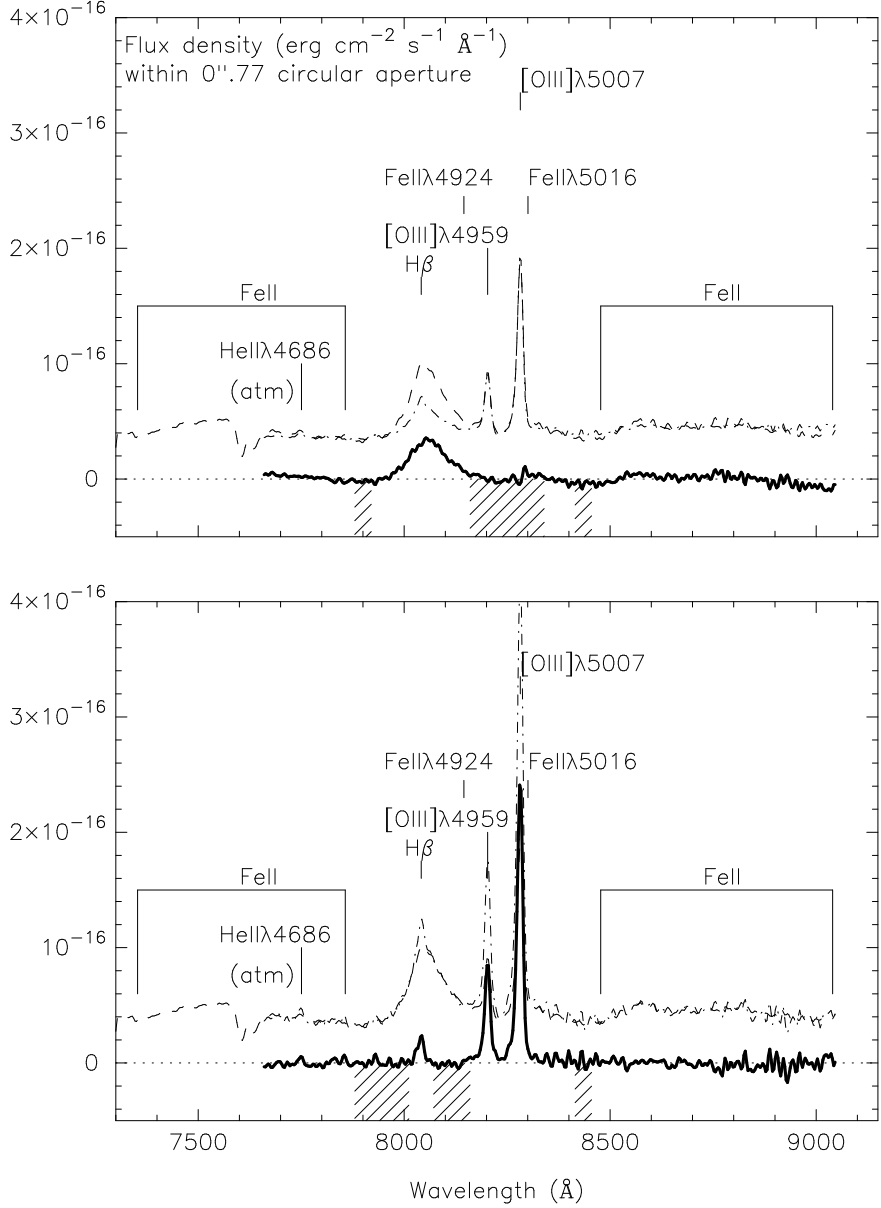}
\caption{
{\it (upper)}
The spectrum of image B (dashed line) has been fit with
the following equation:
$f_B(\lambda) = c_0 \times f_C(\lambda) -
c_1 \times (\lambda/8000 \AA)^{c_2}$,
where $f_B(\lambda)$ and $f_C(\lambda)$ are the flux densities
of images B and C, respectively,
and $c_0$, $c_1$, and $c_2$ are the fitting parameters.
The wavelength regions used for the fitting,
$7880$-$7920 \AA$, $8160$-$8340 \AA$,
and $8415$-$8455 \AA$, are shown with shadows.
These correspond to
the [OIII] and the line-free continuum emission
wavelength regions.
The dot-dashed line is the best fit
($c_0=0.84$, $c_1=-1.57 \times 10^{-17}$, and $c_2=4.2$),
while the thick solid line shows the residuals,
$f_B(\lambda) -$(best fit).
The value of $c_2$ is rather meaningless.
{\it (lower)}
The same as
the above,
except
the wavelength regions used for the fitting
and the resultant fitting parameters.
The wavelength regions are
$7880$-$8011 \AA$, $8071$-$8160 \AA$,
and $8415$-$8455 \AA$,
which correspond to
the broad H$\beta$ and the line-free continuum emission
wavelength regions.
The best fit parameters are
$c_0=2.15$, $c_1=1.79 \times 10^{-17}$, and $c_2=-4.7$.
The thick solid line shows the residuals,
(best fit) $- f_B(\lambda)$.
\label{bcoiiihbbroad}
}
\end{figure}

\begin{figure}
\epsscale{1.0}
\plotone{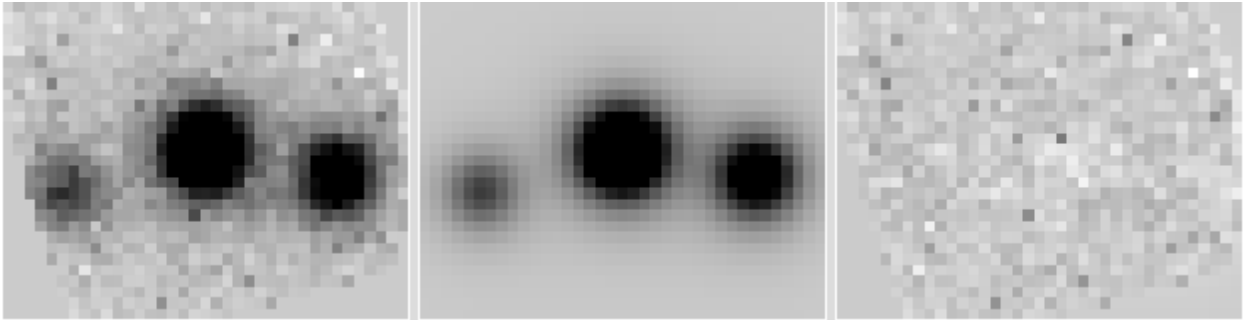}
\caption{
{\it (left)}
Continuum-subtracted H$\beta$(+FeII$\lambda$4924)
line image.
The same as that in Figure~\ref{lineimage}.
{\it (middle)}
Model image with three point sources that
have flux ratios shown in Table~\ref{tbl-ifsspect}
and have the common Moffat function profile.
{\it (right)}
The residual image ((left)-(middle)).
These images are shown in the same grayscale.
\label{psfmoffat}
}
\end{figure}

\begin{figure}
\epsscale{0.333}
\plotone{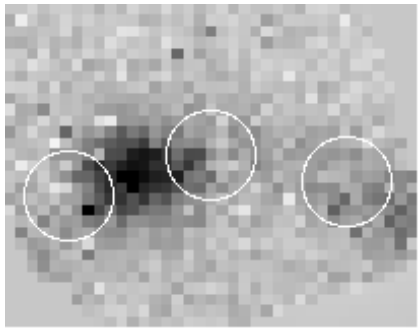}
\caption{
Three-point-source-model subtracted 
[OIII]$\lambda$5007 line image,
where the three point sources
have flux ratios shown in Table~\ref{tbl-ifsspect}
and have the common Moffat function profile as determined
from the broad H$\beta$ point source.
The apertures shown with circles are the same as those in
Figure~\ref{lineimage}.
\label{oiiiextension}
}
\end{figure}

\begin{figure}
\epsscale{1.0}
\plotone{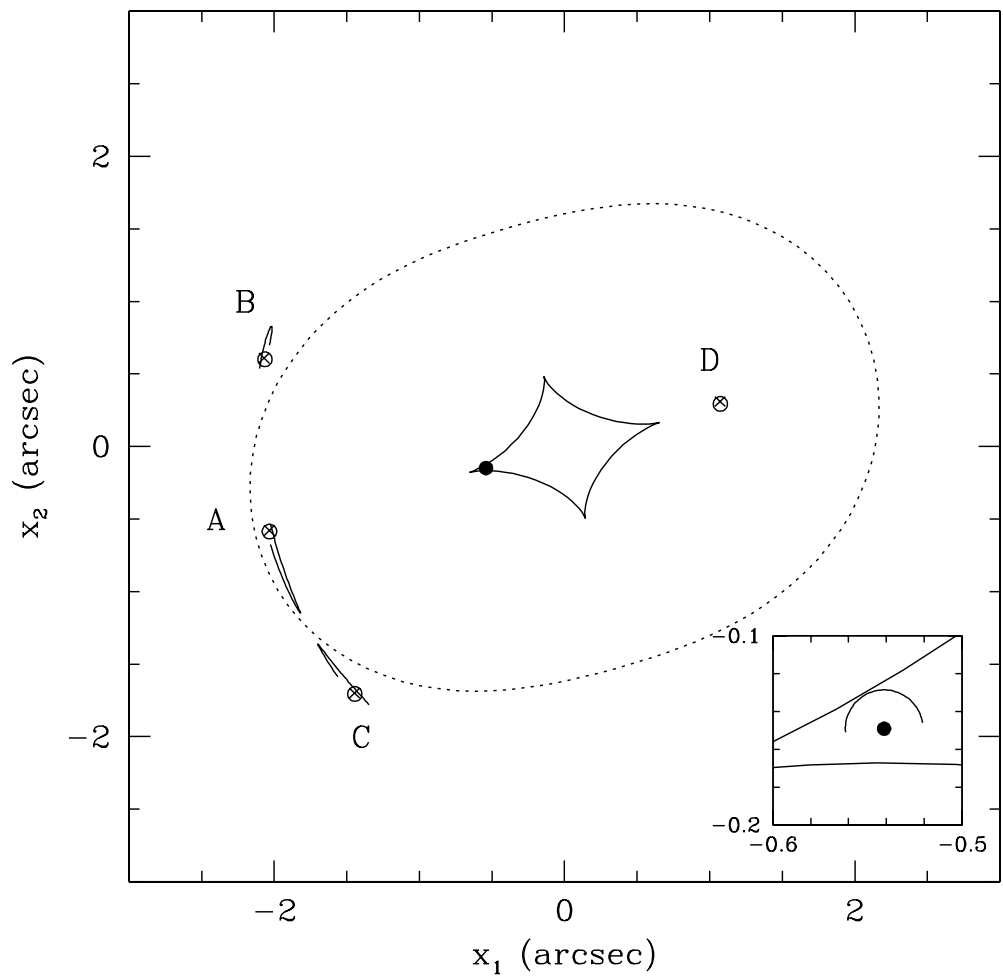}
\caption{
The lens configuration of 1RXS J1131-1231. Open circles show the observed
positions of the images, whereas crosses show the best-fit positions of
the images in our lens model. Solid astroid and dotted ellipse denote
the caustics and critical curve, respectively.
The center of the source position is indicated with a solid circle.
In the inset, the region of the source position is enlarged,
where a surrounding upper semicircle having a radius of 90~pc corresponds to
an asymmetric structure of the NLR in the north-south direction.
This extended, asymmetric source image produces asymmetric arc-like features
(solid curves) near images A, B, and C.
\label{oiiiextensionmodel}
}
\end{figure}

It is worth noting that the observed flux ratio C$/$B in the [OIII] line
emission slightly differs from the model predictions at about the 20 \% level,
while the flux ratio A$/$B is well reproduced.
This slight inconsistency with a smooth-lens model might be caused by
a substructure with an extended instead of point-like mass distribution being 
located at one of the lensed images.
\citet{ino05} showed that for a substructure with a singular
isothermal density distribution, the flux ratio can be changed by about 20~\%
from a smooth-lens prediction even if
$\xi \ga 10$, provided the substructure is just centered at the source image.
Alternatively, and perhaps more likely, is the effect of an asymmetric structure 
or light distribution intrinsic to the NLR \citep[e.g.,][]{schmitt03}.
In contrast, a smooth-lens model assumes a uniform and circularly-symmetric NLR
as a source image. In fact, this effect can partially be seen as an extension of
the [OIII] emission between images A and C as well as a slight extension of
image B in the direction opposite to images A and C (Figure~\ref{lineimage}).
This indicates that we have spatially resolved the NLR itself.
Figure~\ref{oiiiextension} emphasizes these extensions in the subtraction
of three ^^ ^^ point" source models from the [OIII] line image.
The total [OIII] flux of the extended emission between images A and C
is about a half of that of the point-like component of image A
at our resolution.
The fractional contributions from the extended emission components are
16 \% -- 20 \%
even in the $0^{\prime\prime}.77$ apertures of the quasar images.
This causes the slight aperture dependence of [OIII]
flux ratios between images 
(particularly the C$/$B ratio)
as described in \S~\ref{uncertainty}.
We have found, based on the lens modeling, that
asymmetric configurations
of lensed images can be caused by asymmetric structure in the NLR in the
north-south direction, i.e. if the north side of the NLR is more luminous
or more spatially extended than the south side.
An example of such models is shown in Figure~\ref{oiiiextensionmodel}.
This asymmetry in the lensed images of the NLR would be attributed to
a slight difference between the measured flux ratios inside a finite
aperture and the smooth-lens predictions based on a uniform and
circularly-symmetric source.

In contrast to the [OIII] line emission, the broad-line H$\beta$ emission,
which originates from the BLR, shows a significant difference to the model for the
flux ratio C$/$B , i.e. C$/$B$=0.46$ compared with
the predicted ratio of $0.91 \sim 1.00$, while the flux ratio A$/$B is well
reproduced by a smooth lens. This anomaly for the flux ratio C$/$B in the
H$\beta$ line emission is caused neither by effects of (i) dust extinction nor
(ii) time delay on lensed images, as argued below.
(i) There are no effects of dust extinction
on the H$\beta$ line emission, because they are not seen in the
[OIII]$\lambda \lambda$ 4959,5007 lines, whose wavelengths are close to
that of H$\beta$. 
(ii) The time delay between lensed images B and C is a small fraction of a day
as estimated by our smooth-lens model as well as
by \citetalias{slu03} and \citetalias{bla06}.
This is too short for the BLR to change its line flux by a factor of two,
because reverberation mapping \citep[e.g.,][]{pet93} indicates the sizes of BLRs 
in AGNs to be 10$^{1-2}$ light days.
For these reasons, the most likely explanation for the anomalous flux ratio
C$/$B in the H$\beta$ line emission is micro/milli-lensing of image C.
Microlensing of image C would be of particular importance, because of
the finite probability of de-amplification even though it is located at a minimum
in the arrival time surface (Schechter \& Wambsganss 2002, see also
Keeton 2003 for the case of millilensing).
We also note that simultaneous amplification of both images A and B by
similar amounts is another possibility, but it appears to be unlikely
because of the lack of such time variation in their continuum
emission \citepalias{mor06}.

In addition to micro/milli-lensing of image C,
microlensing of image A has occurred such that its
CR is microlensed, while its BLR is only partially affected.
\citetalias{mor06} have shown that the brightness of image A
in optical bands has been increasing over recent years, probably because
of recovery from a de-amplified stage due to microlensing.
Image A was once fainter than image B 
\citepalias{slu03}. Using Table 1 of \citetalias{mor06}, 
we have found that the $R$-band flux ratios
at the time of our observations (HJD$=$2453411) read
A$/$B$\simeq 1.2$ and C$/$B$\simeq 0.5$, implying that the CR for
image A is microlensed. As shown in \S~\ref{compab}, 
an asymmetric profile of
the H$\beta$ line for image A, i.e., weaker blue wing than red wing,
may also indicate the presence
of microlensing for the BLR as well but only in part; such an asymmetric profile
can be caused by the microlensing of the part of the BLR that is rotating
away from us. Thus, the Einstein radius of a substructure (most probably
a star) near image A must be small compared with the size of the BLR.
In contrast, a substructure located near image C would have to be more massive than
that near image A, because it would have to affect the emission from both the CR
and BLR of image C, since the flux ratios C$/$B in both regions
($\sim 0.5$) are almost the same.

The likely radius of the BLR, $R_{\rm BLR}$, is estimated as about
$0.01$ - $0.05$ pc, 
using the relation between $R_{\rm BLR}$
and the intrinsic luminosity of H$\beta$, $L({\rm H}\beta)$,
obtained by \citet{kas05}. For $L({\rm H}\beta)$, we estimate
$L({\rm H}\beta) \sim (1.8$-$4.3) \times 10^{42}$ erg~s$^{-1}$
as derived from the observed H$\beta$ flux
(which is corrected for the limited aperture size
and includes the uncertainty of the absolute flux calibration)
after taking into account
(i) the dust extinction in the lensing galaxy and
(ii) the lensing magnification predicted by lens models,
as detailed below.
(i) Our estimate of dust extinction in the lensing galaxy, which is
an elliptical as deduced from the presence of typical absorption
lines \citepalias{slu03}, is based on work by \citet{eli06}.
They derived, using their sample of mostly early-type lenses,
that the mean differential extinction for the most extinguished image pair
for each lens is $A_V \sim 0.56$ mag ($R_V \sim 2.8$).
We thus adopt $A_V \sim 0$ - $0.56$ mag as a possible range of
the absolute extinction, which corresponds to
$A_{{\rm 0.62}\mu{\rm m}} \sim 0$ - $0.48$ mag in the rest frame
of the lensing galaxy.
(ii) For the lensing magnification, we adopt the predictions of
smooth-lens models, 
yielding 12.4 (this work) to 14.7 \citep{slu03}
for the magnification factor of image B.
Apparently, $L({\rm H}\beta)$ estimated in this manner is subject to
some systematics and the derived $R_{\rm BLR}$ also includes uncertainties
associated with an intrinsic scatter in the $R_{\rm BLR}$ vs. $L({\rm H}\beta)$
relation \citep{kas05}.
Nonetheless, the range of the estimated $R_{\rm BLR}$ of 1RXS J1131-1231,
$0.01$ - $0.05$ pc, 
seems to be comparable to that of a luminous
Seyfert 1 rather than the typical size in luminous quasars ($\sim 0.1$ pc
taken from Fig. 2 of Kaspi et al. 2005). This is actually consistent with
the work by \citetalias{slu03} who, based on the estimation of its unlensed absolute
magnitude, suggested that the source object in the
1RXS J1131-1231
system is a Seyfert~1.

Based on our estimate of 
$R_{\rm BLR} = 0.01$ - $0.05$ pc,
the condition $\xi \la 10$ for the substructure lensing of image C's H$\beta$
line emission requires $M_{\rm E} \ga 0.1$ $M_\odot$ for the mass of
a substructure near image C. On the other hand, for a substructure
near image A, $M_{\rm E}$ must be as small as 0.01 - 0.1 $M_\odot$.
If the mass were larger, the H$\beta$ flux of image A would significantly differ
from the smooth-lens predictions, which is not
the case (see Table~\ref{tbl-ifsspect}).

In summary, we have observed the quadruply lensed quasar 1RXS J1131-1231
with the IFS mode of the Kyoto 3DII.
The simultaneous observation of the three brighter quasar
images at high spatial resolution has enabled us to measure
 accurate flux ratios between them
in the H$\beta$ and [OIII] lines,
and to clarify the cause of anomalous flux ratios
also seen in the continuum emission.
We have found that the flux ratios in the NLR [OIII] line
can be explained by smooth-lens models, with only a slight
deviation from the model predictions.
The deviation is likely caused by 
asymmetric structure in the NLR.
The absence of microlensing and/or millilensing effects on the [OIII]
line emission sets a tight limit of $M_{\rm E} < 10^5 M_\odot$
along the light path to the NLR.
The BLR H$\beta$ shows a C$/$B line flux ratio that is smaller than the
smooth-lens model predictions.
This is most likely caused by the micro/milli-lensing of image C.
The slight difference of the broad H$\beta$ line profile in
image A compared with those in the other images suggests
the presence of a small microlensing effect on image A.
Our results demonstrate that the IFS observations of
gravitational lens systems are useful
for understanding the mass distribution of lensing galaxies
and the internal structure of lensed sources.

\acknowledgments

We thank the staff at Subaru Telescope for their help during
the observing run.
We acknowledge partial financial support from NAOJ,
including Subaru Telescope.
We also thank
S. Ozaki, H. Ohtani, T. Hayashi, T. Ishigaki, M. Ishii, M. Sasaki,
Y. Okita, T. Minezaki, and R. I. Davies for discussions,
and the referee for helpful comments.
This work is supported by the Grant-in-Aid for the 21st Century COE
^^ ^^ Center for Diversity and Universality in Physics" from the
Ministry of Education, Culture, Sports, Science and Technology
(MEXT) of Japan.

%% To help institutions obtain information on the effectiveness of their
%% telescopes, the AAS Journals has created a group of keywords for telescope
%% facilities. A common set of keywords will make these types of searches
%% significantly easier and more accurate. In addition, they will also be
%% useful in linking papers together which utilize the same telescopes
%% within the framework of the National Virtual Observatory.
%% See the AASTeX Web site at http://www.journals.uchicago.edu/AAS/AASTeX
%% for information on obtaining the facility keywords.

%% After the acknowledgments section, use the following syntax and the
%% \facility{} macro to list the keywords of facilities used in the research
%% for the paper.  Each keyword will be checked against the master list during
%% copy editing.  Individual instruments can be provided in parentheses,
%% after the keyword, but they will not be verified.

Facilities: \facility{Subaru(Kyoto 3DII)}.

\clearpage

\begin{table}
\begin{center}
\caption{Relative flux ratios among quasar images
A, B, and C.\label{tbl-ifsspect}}
\begin{tabular}{cccc}
\tableline\tableline
Line\tablenotemark{a} & A & B & C \\
\tableline
[OIII]                     &$1.63^{+0.04}_{-0.02}$ &1.00   &$1.19^{+0.03}_{-0.12}$\\
H$\beta$ (broad)           &$1.74^{+0.07}_{-0.12}$ &1.00   &$0.46^{+0.02}_{-0.03}$\\
\tableline
SIEx$+$\tablenotemark{b} &1.66 &1.00 &0.91 \\
SISx\tablenotemark{c} &1.75 &1.00 &1.00 \\
SISx\tablenotemark{d} &1.70 &1.00 &0.96 \\
\tableline
\end{tabular}

\tablenotetext{a}{Observed on 2005 February 8.
Determined from the scaling and fitting procedures  
between quasar image spectra
by using an 8-lenslet ($0^{\prime\prime}.77$) diameter
pure
circular aperture each.
See text for the uncertainties of flux ratios.
}
\tablenotetext{b}{Our smooth-lens model
with a singular
isothermal ellipsoid plus an external shear, 
based on the quasar image positions obtained by \citetalias{mor06}.
A slight offset 
of 9 milliarcseconds 
has been applied for the halo center with respect to the 
lensing 
galaxy center.}
\tablenotetext{c}{Singular
isothermal sphere plus an external shear model
by \citetalias{slu03}.}
\tablenotetext{d}{Singular
isothermal sphere plus an external shear model
by \citetalias{bla06}.}
\end{center}
\end{table}

%% Use the figure environment and \plotone or \plottwo to include
%% figures and captions in your electronic submission.
%% To embed the sample graphics in
%% the file, uncomment the \plotone, \plottwo, and
%% \includegraphics commands
%%
%% If you need a layout that cannot be achieved with \plotone or
%% \plottwo, you can invoke the graphicx package directly with the
%% \includegraphics command or use \plotfiddle. For more information,
%% please see the tutorial on "Using Electronic Art with AASTeX" in the
%% documentation section at the AASTeX Web site,
%% http://www.journals.uchicago.edu/AAS/AASTeX.
%%
%% The examples below also include sample markup for submission of
%% supplemental electronic materials. As always, be sure to check
%% the instructions to authors for the journal you are submitting to
%% for specific submissions guidelines as they vary from
%% journal to journal.

%% This example uses \plotone to include an EPS file scaled to
%% 80% of its natural size with \epsscale. Its caption
%% has been written to indicate that additional figure parts will be
%% available in the electronic journal.

%% Here we use \plottwo to present two versions of the same figure,
%% one in black and white for print the other in RGB color
%% for online presentation. Note that the caption indicates
%% that a color version of the figure will be available online.
%%

%% This figure uses \includegraphics to scale and rotate the still frame
%% for an mpeg animation.
\clearpage

\clearpage
\pagestyle{empty}
\setlength{\voffset}{-25mm}

\clearpage
\setlength{\voffset}{-15mm}

\clearpage
\pagestyle{plaintop}
\setlength{\voffset}{0mm}

%% If you are not including electonic art with your submission, you may
%% mark up your captions using the \figcaption command. See the
%% User Guide for details.
%%
%% No more than seven \figcaption commands are allowed per page,
%% so if you have more than seven captions, insert a \clearpage
%% after every seventh one.

%% Tables should be submitted one per page, so put a \clearpage before
%% each one.

%% Two options are available to the author for producing tables:  the
%% deluxetable environment provided by the AASTeX package or the LaTeX
%% table environment.  Use of deluxetable is preferred.
%%

%% Three table samples follow, two marked up in the deluxetable environment,
%% one marked up as a LaTeX table.

%% In this first example, note that the \tabletypesize{}
%% command has been used to reduce the font size of the table.
%% We also use the \rotate command to rotate the table to
%% landscape orientation since it is very wide even at the
%% reduced font size.
%%
%% Note also that the \label command needs to be placed
%% inside the \tablecaption.

%% This table also includes a table comment indicating that the full
%% version will be available in machine-readable format in the electronic
%% edition.
%%


\begin{thebibliography}{34}
\expandafter\ifx\csname natexlab\endcsname\relax\def\natexlab#1{#1}\fi

\bibitem[{Blackburne {et~al.}(2006)Blackburne, Pooley, \& Rappaport}]{bla06}
Blackburne, J.~A., Pooley, D., \& Rappaport, S. 2006, \apj, 640, 569 (BPR06)

\bibitem[{Chiba(2002)}]{chi02}
Chiba, M. 2002, \apj, 565, 17

\bibitem[{Chiba {et~al.}(2005)Chiba, Minezaki, Kashikawa, Kataza, \&
  Inoue}]{chi05}
Chiba, M., Minezaki, T., Kashikawa, N., Kataza, H., \& Inoue, K.~T. 2005, \apj,
  627, 53

\bibitem[{Claeskens {et~al.}(2006)Claeskens, Sluse, Riaud, \& Surdej}]{cla06}
Claeskens, J.-F., Sluse, D., Riaud, P., \& Surdej, J. 2006, \aap, 451, 865

\bibitem[{Dalal \& Kochanek(2002)}]{dal02}
Dalal, N. \& Kochanek, C.~S. 2002, \apj, 572, 25

\bibitem[{El\'iasd\'ottir {et~al.}(2006)El\'iasd\'ottir, Hjorth, Toft, Burud,
  \& Paraficz}]{eli06}
El\'iasd\'ottir, A., Hjorth, J., Toft, S., Burud, I., \& Paraficz, D. 2006,
  \apjs, 166, 443

\bibitem[{Inoue \& Chiba(2005)}]{ino05}
Inoue, K.~T. \& Chiba, M. 2005, \apj, 634, 77

\bibitem[{Iye {et~al.}(2004)}]{iye04}
Iye, M. {et~al.} 2004, \pasj, 56, 381

\bibitem[{Kaspi {et~al.}(2005)Kaspi, Maoz, Netzer, Peterson, Vestergaard, \&
  Jannuzi}]{kas05}
Kaspi, S., Maoz, D., Netzer, H., Peterson, B.~M., Vestergaard, M., \& Jannuzi,
  B.~T. 2005, \apj, 629, 61

\bibitem[{Kassiola \& Kovner(1993)}]{kas93}
Kassiola, A. \& Kovner, I. 1993, \apj, 417, 450

\bibitem[{Keeton(2003)}]{kee03}
Keeton, C.~R. 2003, \apj, 584, 664

\bibitem[{Klypin {et~al.}(1999)Klypin, Kravtsov, Valenzuela, \& Prada}]{kly99}
Klypin, A., Kravtsov, A.~V., Valenzuela, O., \& Prada, F. 1999, \apj, 522, 82

\bibitem[{Kochanek {et~al.}(2006)Kochanek, Schneider, \& Wambsganss}]{koc06}
Kochanek, C., Schneider, P., \& Wambsganss, J. 2006, in Gravitational Lensing:
  Strong, Weak \& Micro, ed. G.~Meylan, P.~Jetzer, \& P.~North, Proceedings of
  the 33rd Saas-Fee Advanced Course (Heidelberg: Springer-Verlag)

\bibitem[{Kormann {et~al.}(1994)Kormann, Schneider, \& Bartelmann}]{kor94}
Kormann, R., Schneider, P., \& Bartelmann, M. 1994, \aap, 284, 285

\bibitem[{Mao \& Schneider(1998)}]{mao98}
Mao, S. \& Schneider, P. 1998, \mnras, 295, 587

\bibitem[{Metcalf \& Madau(2001)}]{met01}
Metcalf, R.~B. \& Madau, P. 2001, \apj, 563, 9

\bibitem[{Metcalf {et~al.}(2004)Metcalf, Moustakas, Bunker, \& Parry}]{met04}
Metcalf, R.~B., Moustakas, L.~A., Bunker, A.~J., \& Parry, I.~R. 2004, \apj,
  607, 43

\bibitem[{Moore {et~al.}(1999)Moore, Ghigna, Governato, Lake, Quinn, Stadel, \&
  Tozzi}]{moo99}
Moore, B., Ghigna, S., Governato, F., Lake, G., Quinn, T., Stadel, J., \&
  Tozzi, P. 1999, \apj, 524, L19

\bibitem[{Morgan {et~al.}(2006)Morgan, Kochanek, Falco, \& Dai}]{mor06}
Morgan, N.~D., Kochanek, C.~S., Falco, E.~E., \& Dai, X. 2006, astro-ph/0605321
  (M06)

\bibitem[{Moustakas \& Metcalf(2003)}]{mou03}
Moustakas, L.~A. \& Metcalf, R.~B. 2003, \mnras, 339, 607

\bibitem[{Peterson(1993)}]{pet93}
Peterson, B.~M. 1993, \pasp, 105, 247

\bibitem[{Schechter \& Wambsganss(2002)}]{sch02}
Schechter, P.~L. \& Wambsganss, J. 2002, \apj, 580, 685

\bibitem[{Schmitt {et~al.}(2003)Schmitt, Donley, Antonucci, Hutchings, \&
  Kinney}]{schmitt03}
Schmitt, H.~R., Donley, J.~L., Antonucci, R. R.~J., Hutchings, J.~B., \&
  Kinney, A.~L. 2003, \apjs, 148, 327

\bibitem[{Sluse {et~al.}(2006)Sluse, Claeskens, Altieri, Cabanac, Garcet,
  Hutsem\'ekers, Jean, Smette, \& Surdej}]{slu06}
Sluse, D., Claeskens, J.-F., Altieri, B., Cabanac, R.~A., Garcet, O.,
  Hutsem\'ekers, D., Jean, C., Smette, A., \& Surdej, J. 2006, \aap, 449, 539

\bibitem[{Sluse {et~al.}(2003)}]{slu03}
Sluse, D. {et~al.} 2003, \aap, 406, L43 (S03)

\bibitem[{Sugai {et~al.}(2006)Sugai, Kawai, Hattori, Ozaki, Kosugi, Shimono, \&
  Okita}]{sug06}
Sugai, H., Kawai, A., Hattori, T., Ozaki, S., Kosugi, G., Shimono, A., \&
  Okita, Y. 2006, New Astronomy Reviews, 50, 358

\bibitem[{Sugai {et~al.}(2002)Sugai, Ozaki, Hattori, \& Kawai}]{sug02}
Sugai, H., Ozaki, S., Hattori, T., \& Kawai, A. 2002, in Astron. Soc. Pacif.
  Conf. Ser., Vol. 282, Galaxies: The Third Dimension, ed. M.~Rosado,
  L.~Binette, \& L.~Arias (San Francisco: Astronomical Society of the Pacific),
  433

\bibitem[{Sugai {et~al.}(2000)}]{sug00b}
Sugai, H. {et~al.} 2000, \procspie, 4008, 558

\bibitem[{Sugai {et~al.}(2004)}]{sug04}
---. 2004, \procspie, 5492, 651

\bibitem[{Sugai {et~al.}(2005)}]{sug05}
---. 2005, \apj, 629, 131

\bibitem[{V\'eron-Cetty {et~al.}(2004)V\'eron-Cetty, Joly, \& V\'eron}]{ver04}
V\'eron-Cetty, M.-P., Joly, M., \& V\'eron, P. 2004, \aap, 417, 515

\bibitem[{Wayth {et~al.}(2005)Wayth, O'Dowd, \& Webster}]{way05}
Wayth, R.~B., O'Dowd, M., \& Webster, R.~L. 2005, \mnras, 359, 561

\bibitem[{Wills {et~al.}(1993)Wills, Brotherton, Fang, Steidel, \&
  Sargent}]{wil93}
Wills, B.~J., Brotherton, M.~S., Fang, D., Steidel, C.~C., \& Sargent, W. L.~W.
  1993, \apj, 415, 563

\bibitem[{Wyithe {et~al.}(2002)Wyithe, Agol, \& Fluke}]{wyi02}
Wyithe, J. S.~B., Agol, E., \& Fluke, C.~J. 2002, \mnras, 331, 1041

\end{thebibliography}
\end{document}